\documentclass[12pt]{revtex4}
\usepackage{graphicx}
\usepackage{amstext}
\usepackage{amssymb, amsmath}
\usepackage{amsthm}
\usepackage{natbib}

\begin{document}
\title{Separating electric field and thermal effects across the metal-insulator transition in vanadium oxide nanobeams}

\author{Adam A. Stabile$^{1}$}
\author{Sujay K. Singh$^{1}$}
\author{Tai-Lung Wu$^{1}$}
\author{Luisa Whittaker$^{2}$}
\author{Sarbajit Banerjee$^{2}$}
\author{G. Sambandamurthy$^{1}$}
\email{sg82@buffalo.edu}
\affiliation{$^{1}$Department of Physics}
\affiliation{$^{2}$Department of Chemistry, University at Buffalo, State University of New York, Buffalo, New York 14260, USA}

\begin{abstract}

We present results from an experimental study of the equilibrium and non-equilibrium transport properties of vanadium oxide nanobeams near the metal-insulator transition (MIT). Application of a large electric field in the insulating phase across the nanobeams produces an abrupt MIT and the individual roles of thermal and non-thermal effects in driving the transition are studied. Transport measurements at temperatures ($T$) far below the critical temperature ($T_c$) of MIT, in several nanoscale vanadium oxide devices, show that both $T$ and electric field play distinctly separate, but critical roles in inducing the MIT. Specifically, at $T << T_c$ electric field dominates the MIT through an avalanche-type process, whereas thermal effects become progressively critical as $T$ approaches $T_c$. 

\end{abstract}

\maketitle

In strongly correlated Mott insulating systems wherein electron-electron Coulomb interactions induce a deviation from behavior expected from simple band theory considerations, the ability to modulate charge carrier density provides unique opportunities to explore new phases and transitions between them  \cite{Dagotto05,Ahn03,Morosan12}. Moreover, in oxide materials  exhibiting  pronounced, thermally driven metal-insulator transitions (MIT), the application of an electric field in the insulating state also results in a MIT  \cite{Kim04,Lee08}. However, thermal effects inherent in such measurements have raised questions regarding the mechanisms underpinning the MIT, a debate that still rages to this day \cite{Lee08,Zim13,Yang11}. In this Letter, we provide original analyses of the ways in which electric field and temperature affect the MIT in single-crystal nanobeams of substitutionally doped (with W) VO$_2$. In doing so, we conclusively show that electric field is a unique parameter, distinct from temperature, for induction of the dramatic phase transition in vanadium oxide nanobeams.

The hallmark property of VO$_2$ is the MIT: when crossing a critical temperature, $T_c \backsim$ 340 K, a dramatic switch in electrical resistance occurs over several orders of magnitude \cite{Morin59} and the MIT is also accompanied by a structural phase transition (SPT): the unit cell changes from a monoclinic M1 phase (insulating phase) to a rutile R phase (metallic phase) \cite{Eyert11,Mott68,Morin59}. Recent work on nanostructures of VO$_2$ has revealed another interesting phenomenon: depending on certain extrinsic parameters (such as stress or sample growth conditions) the MIT may occur not as a sharp switch between M1 and R phases, but in discrete steps wherein domains  of M1 and R phases coexist at temperatures around $T_c$ \cite{Qazil07,Wei09,Zhang09}. Such an intriguing phase coexistence has been seen in several  systems, and such systems provide a rich landscape for exploring unique correlated electron phenomena \cite{Dagotto05,Qazil07}.  

Although the MIT  in VO$_2$ has been explored for more than half of a century \cite{Morin59}, the solid state community continues to look for a direct answer to the carrier-mediated mechanisms that constitute the mechanistic basis of this abrupt transition \cite{Mott68,Eyert11,Qazil07,Park13}. Although it might seem ideal to study the influence of electric fields on the properties of VO$_2$ systems using gated devices, these measurements have been marred by seemingly intractable problems. Continuous carrier injection using gated devices with oxide dielectrics such as SiO$_2$ or high-$k$ materials have resulted in unwanted interface effects or do not result in adequate carrier injection required to induce the MIT \cite{Stefano00,Sen11}. Recently, ionic liquid gating has shown promise in emerging as a methodology for such electric field studies, but effects from hydrogen doping or oxygen vacancies have been shown to alter the atomic structure of oxide materials, including VO$_2$ \cite{Nakano12,Wei12,Yang12,Jeong13}. 

Another approach to investigating this problem is by studying the non-equilibrium application of an electric field applied between the source and drain of two-terminal oxide devices \cite{Kim04,Lee08,Zim13,Yang11,Stefano00}. Several groups have observed such an electric field-driven MIT (E-MIT) through the observation of an orders of magnitude switch in current across a certain threshold voltage in vanadium oxide systems \cite{Kim04,Wu11,Wu11_2} . Many researchers believe that the $\textit{product}$ of the current and voltage (Joule heating), and not voltage alone (electric field), is the critical parameter driving the MIT in this type of measurement \cite{Zim13,Mun13}. Voltage pulsed studies minimize Joule heating effects by measuring phenomena in time scales outside the realm of thermal dissipation. Although such measurements have proven successful in establishing the MIT in magnetite Fe$_3$O$_4$ as an electric field induced phenomenon \cite{Lee08}, measurements performed on VO$_2$ have yet to conclusively show distinct characteristics of an E-MIT \cite{Chae05,Zim13} 

In this context, our work takes a novel approach to the non-equilibrium electric field measurements on VO$_2$ systems by quantitatively measuring the effects both from the electric field as well as from the accompanying thermal effects. We present three main results that clearly show that at temperatures far below the MIT, a MIT can be induced purely from electric field effects and thermal effects progressively dominate the transition as $T_c$ is approached. 
All measurements are performed using multi-terminal devices made from free standing, single-crystal nanobeams of W-doped VO$_2$ \cite{Wu11,Luisa11_2}. The nanobeams are rectangular in cross section with typical dimensions of 100 nm (thickness) X 500 nm (width). 
They are substitutionally doped with tungsten (W$_x$V$_{1-x}$O$_2$ where $\textit{x}$ = 0.6243 \% is the atomic percentage of W). The tungsten doping is known to reduce  $T_c$ to around room temperature making W-VO$_2$ particularly attractive from a commercial application standpoint \cite{Chae10}. Moreover, these W-VO$_2$ nanobeams have proven successful in exploring and in providing new insights into the elusive properties of VO$_2$ systems without loss of generality, and without stabilization of metastable M2 and triclinic phases often observed at the nanoscale as a result of  strain \cite{Luisa11,Park13}.

%

\begin{figure}[h]
\includegraphics[height=5.2in]{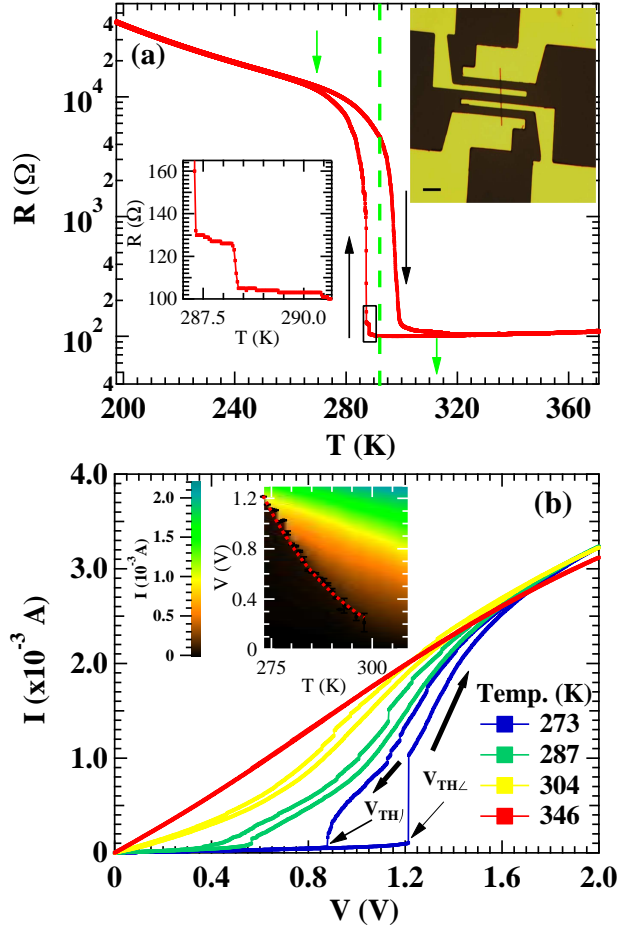}    
\caption{\textbf{Transport characteristics of a single nanobeam.} (a) T-MIT of a single W-VO$_2$ nanobeam measured in 4-terminal configuration in the zero bias limit. Black arrows indicate sweep direction. Dashed green line defines $T_c$ = 291 K as the midpoint between the beginning and the end of the hysteresis (green arrows). Bottom inset: $R$ vs. $T$ within the data range of boxed outline. Top inset: Optical image of a single nanobeam device. Scale bar = 20 $\mu$m. (b) E-MIT of a single W-VO$_2$ nanobeam ($T_c$ = 291 K). Bold arrows indicate sweep direction. Inset: Color plot of current as a function of voltage and temperature where voltage is tuned from 0 to 2 V. Dotted trace is fit to threshold voltage as described in text.} 
\label{FIG.1}
\end{figure}

Methods for synthesizing and fabricating single nanobeam devices (channel lengths = 0.5 $\mu$m or 6 $\mu$m) are described elsewhere [20]. Fig. 1 (a) shows resistance ($R$) as a function of temperature ($T$) for a single nanobeam device of W-VO$_2$ (Fig. 1 (a) top inset) in which we observe a transition between insulator and metal with $T_c$ = 291 K. Within the MIT, we also observe several discontinuous jumps in $R$ (Fig. 1 (a) inset) \cite{Sharoni08}; these jumps are attributed to the formation/nucleation of domains manifested from stresses developed within the nanobeam across the SPT \cite{Cao09}. Similar to the $T$-driven case, we observe a transition between metal and insulator upon the application of a DC voltage, and the plot indicates several  small jumps within the current-voltage ($\textit{IV}$) curve (Fig. 1 (b)) \cite{Kim04,Wu11} These transitions occur at threshold voltages $V_{TH\uparrow}$ and $V_{TH\downarrow}$ on the up sweep and on the down sweep respectively for $T < T_c$. As temperature increases, the transport characteristics change since threshold voltages become no longer discernible, and the hysteretic width gradually shrinks. The color plot of Fig. 1 (b) inset summarizes the $\textit{IV}$ characteristics of our sample on the up sweep. The dotted red-colored fit traces the temperature dependence of $V_{TH\uparrow}$, and has the form typically seen in charge ordered systems $\backsim$ $e^{-T/T_{o}}$. On the down sweep (not shown), $V_{TH\downarrow}$ follows the Joule heating trend $\backsim$ $\sqrt{T_{c}-T}$. Both trends have been discussed previously \cite{Wu11}.

\begin{figure}[h]
\includegraphics[height=5in]{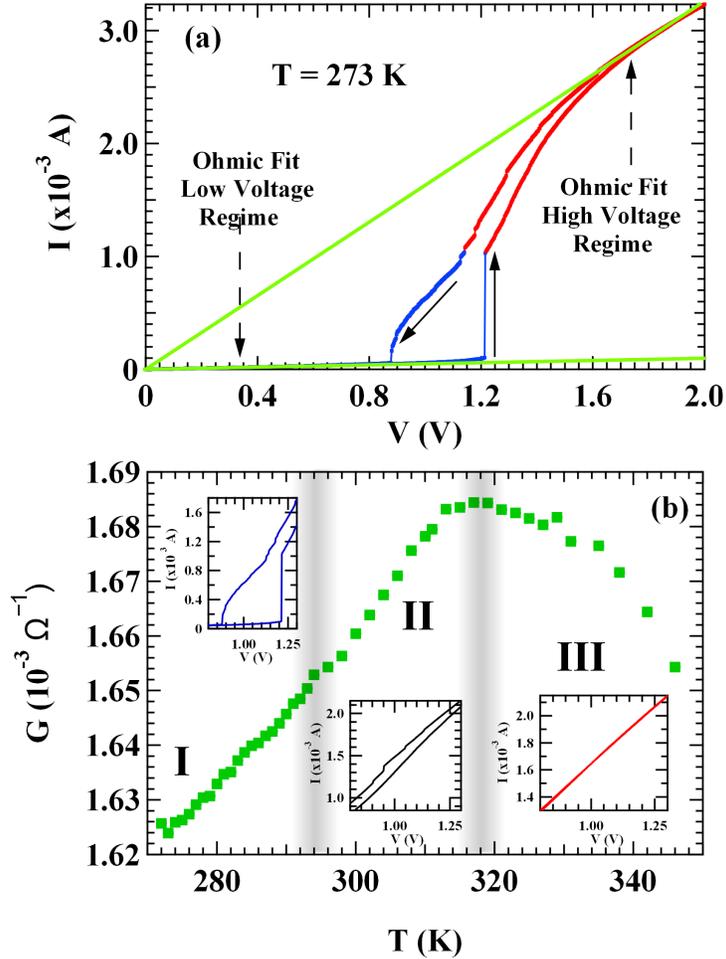}
\caption{\textbf{Strange metal phase.} (a) $\textit{IV}$ curve showing linear fits with zero intercept in the high voltage and in the low voltage regimes. Meaning of different colored sections described in text. (b) Conductance ($\textit{G}$) obtained from linear fits in the high voltage regime as a function of $T$. Regions I, II, and III qualitatively delineate three distinct shapes of the $\textit{IV}$ curves as shown in the insets.  Insets: $\textit{IV}$ curves at temperatures 273 K, 303 K and 346 K. Note that $T_{c}$ = 291 K borders region I and II. }
\label{FIG.2}
\end{figure}

We start our analyses by carefully examining the linearity of the $IV$ curves. In a typical $\textit{IV}$ curve we can interpolate an ohmic fit through the origin in the ranges both before and after the transition (Fig. 2 (a)). Thus, we plot the conductance ($\textit{G}$) (obtained from the linear fits in the high voltage Ohmic regime above $V_{TH}$)  for all values of $T$ in Fig. 2 (b). Since the nanobeam is driven to a metallic state by an electric field, it is expected that the $G$ vs. $T$ plot will be monotonic. Interestingly, we observe a non-monotonic behavior in which a maximum occurs at 318 K which is 27 K above $T_{c}$ for this sample. Above 318 K, typical metallic behavior in $G$ is recovered. However, below 318 K, it appears that the electric field is generating a metallic state, which is endowed with some distinctly non-metallic characteristics. This non-metallic behavior (increase in conductance with $T$) well above $V_{TH}$ is surprising. This suggests that the ways in which electric field and $T$ drive the material to a metallic state are different, and that the E-MIT creates a correlated metallic phase with non-metallic characteristics which persists well beyond $T_{c}$.

\begin{figure}[h]
\begin{center}
\includegraphics[height=5in]{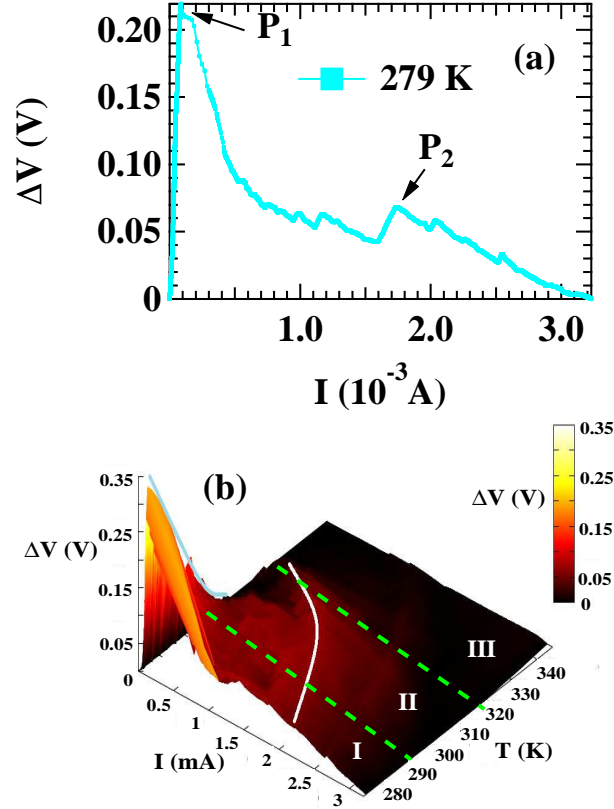}       
\caption{\textbf{Hysteretic width behavior.} (a) Hysteretic width ($\Delta\textit{V}$) vs. $\textit{I}$ at 279 K showing two general peaks $P_1$ and $P_2$. (b) 3D surface plot of $\Delta\textit{V}$ vs. current and temperature. Dotted green lines delineate regions I - III. The two curved lines guide the eye over $P_1$ and $P_2$ showing their temperature and current dependencies.}
\end{center}
\label{FIG.3}
\end{figure} 
It is clear from Fig. 1 (b) that thermal effects dramatically change the overall shape of $\textit{IV}$ curves across the E-MIT. Specifically, the hystereses tend to have three qualitatively distinct shapes at the three regions as shown in Fig. 2 (b): In region I ($T \lesssim T_c$) the shape is non-uniform  and characterized by distinct threshold voltages; in region II ($T \sim T_c$), the shape is more uniform but the threshold voltages begin to vanish; in region III, the hystereses are absent as $T > T_c$ (see inset of Fig. 2(b)). We study these hysteretic shapes and their evolution with $T$ in more detail (Fig. 3) by plotting the hysteretic width ($\Delta\textit{V}$) at every point in the $\textit{IV}$ curves as a function of current for a given $T$. Fig. 3 (a) is a typical trace for $\textit{T} < T_c$ showing two prominent features. The first feature ($P_1$) and second feature ($P_2$) are defined as the maximum hysteretic widths in the blue and in the red sections respectively from Fig. 2 (a). A 3D plot is constructed to study the $T$ evolution of $\Delta\textit{V}$ vs. $I$ traces in Fig. 3(a). From the 3D surface plot of Fig. 3 (b), we observe that the two peaks have distinctly different $T$-dependence in its size (amplitude) and in its position along the current axis. We explore these dependencies in  detail in Fig. 4.

\begin{figure}[h]
\begin{center}
\includegraphics[height=4in]{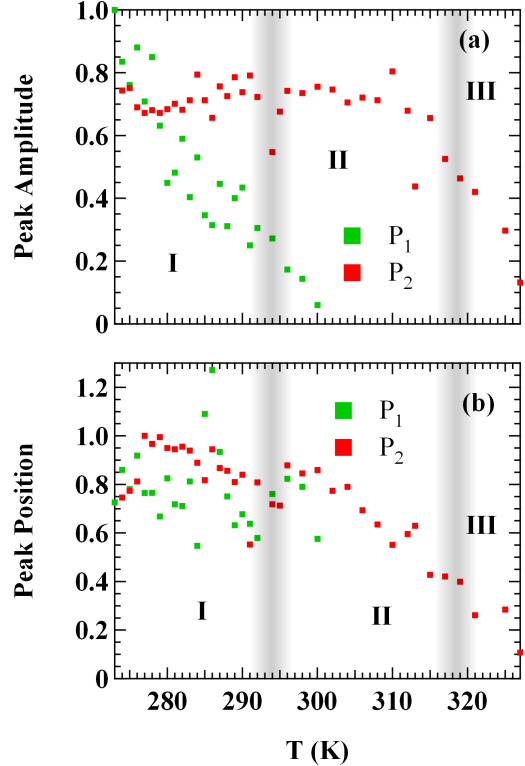}       
\caption{\textbf{Evolution of hysteretic widths.} (a) $\Delta\textit{V}$ scaled to the maximum value of $\Delta\textit{V}$ vs. $T$ for the two prominent features, $P_1$ and $P_2$, defined in Fig. 3(a). (b) $T$-evolution of the positions of $P_1$ and $P_2$ from Fig. 3(b) (values are normalized to the maximum).}
\end{center}
\label{FIG.4}
\end{figure} 
In region I (Fig. 4 (a)), the amplitude of $P_1$ has a strong temperature dependence compared with that of $P_2$. Far into region II (above 300 K) in which $P_1$ is no longer discernible, $P_2$'s amplitude remains almost $T$ independent, but it shows movement along the current axis towards smaller values of $\textit{I}$ (Fig. 4(b)). $P_1$ shows no such dependence in region I. Expectedly, hysteresis is no longer discernible inside region III exemplified by the decrease in $P_2$'s amplitude and position along the current axis (the nanobeam is in a fully metallic state in this regime). The markedly different $T$ dependencies of the amplitudes and of the positions along the current axis of these features underscore distinct mechanisms modulating the shape of the hystereses: electric field dominates the phenomena underlying $P_1$ and temperature dominates the phenomena underlying $P_2$. Even though $V_{TH\downarrow}$ is described by Joule heating effects, the avalanche effect observed at $V_{TH\uparrow}$, especially at temperatures far below $T_c$, is of a different origin (electric field), and has a stronger temperature dependence than that of $V_{TH\downarrow}$; this likely explains the sharp decrease in $P_1$'s amplitude with temperature in region I. However, $P_2$ is probably manifested from thermal-induced effects. After $V_{TH\uparrow}$, Joule heating can become significant: the current jumps by more than a factor of 10 which may raise the temperature of the sample above that of the set temperature \cite{Zim13}. However, the effects of Joule heating become less significant as the set temperature increases. Thus, the decrease in $P_2$'s position with increasing temperature may be a direct reflection of the decreasing prominence of Joule heating in regions II - III. 



\begin{figure}[h]
\begin{center}
\includegraphics[height=2.5in]{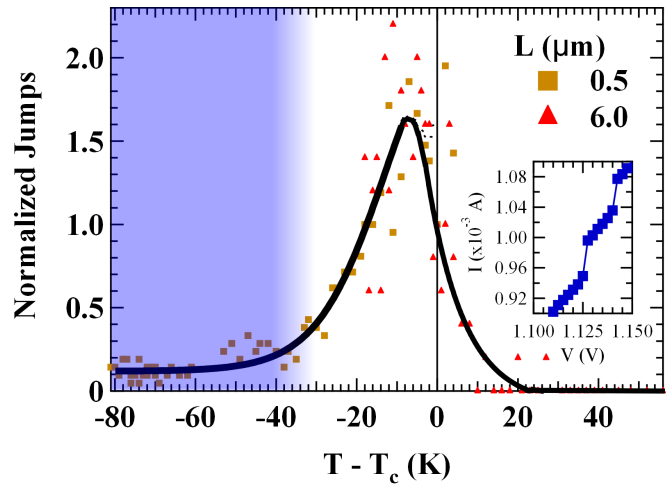}       
\caption{\textbf{Temperature evolution of domains.} Number of jumps over number of jumps at $T_c$ vs. $T - T_c$ measured during the up sweep for two nanobeam devices of different channel lengths, $\textit{L}$. Solid line guides the eye and the colored section separates electric field effects (blue) from thermal effects (white) on domain formation. Inset: Zoomed in section of the 273 K trace from Fig. 1 (b).}
\end{center}
\label{FIG.5}
\end{figure} 

The third and final analysis strives to separate the role of E- and of T-MIT by analyzing their roles in the formation of nanoscale domains. It has been shown that discrete jumps in low bias, $T$-driven measurements of VO$_2$ systems with confined geometry correlate to states of phase separation between metallic and insulating regions \cite{Sharoni08}; however, the role of electric field in forming domains is still not clear. Thus, we plot the number of jumps found from $\textit{IV}$ curves as a function of ($T - T_c$) on the up sweep across the E-MIT (Fig. 5). Jumps are counted through the observation of discontinuous step-like increases in current (Fig. 5 inset). In each  $\textit{IV}$ trace, the magnitude of jumps range from 10 $\mu$A to a few mA, and the number  of jumps range from $\backsim$ 1 - 40 in our devices. Since two single nanobeam devices with different channel lengths ($\textit{L}$ = 0.5 and 6 $\mu$m) are used in this analysis, the number of jumps are scaled to the number of jumps at their respective values of $T_c$ \cite{Sharoni08}. Notably, both nanobeam devices show similar trends:  At temperatures far below $T_c$, the number of jumps is relatively low and constant, and the MIT is often characterized by the single jump at $V_{TH\uparrow}$ (see Fig. 1(b)) indicating dominance of an avalanche-type process induced by the electric field. Hence, thermal effects seem absent in this region and is evident that electric field drives the MIT without the mediating phase coexistence. As $T$ approaches $T_c$, the number of jumps sharply increases and reaches a maximum just around $T_c$. Thus, although threshold voltages are still discernible in this region,thermal-driven effects on the MIT begin to appear by manifesting domains. At $T$ just above $T_c$, phase separation still exists as seen from the jumps, but the system becomes more uniformly metallic and the number of jumps drastically decrease. For $T - T_c \gtrsim$ 20 K, the system is homogeneously metallic, and thus no more jumps are observed. Strain and inhomogeneous doping profiles along the length of an individual nanobeam likely lead to the nucleation of metallic domains across a broad range of temperatures. In past work, we have elucidated the symmetric local structure adopted in proximity of dopant sites, which creates regions structurally analogous to the high-temperature rutile phase, and likely serve as sites for nucleating the structural phase transition \cite{Luisa11_2}. The development of phase boundaries accompanies local nucleation of rutile domains in the otherwise monoclinic nanobeam. Establishment of a percolative pathway across the length of the nanobeam is thus manifested in the transport data as a series of discontinuous step-like events. At low temperatures, a percolative pathway can be established by induction of a parallel electric field without having to overcome distinct phase boundaries and is thus evidenced as an avalanche-type process \cite{Sheka11}. 
The greater abundance of avalanche type (versus discrete step-like) processes at lower temperature and the steady decrease of such processes with increasing temperature further correlates with diminution of the amplitude of P1 with increasing temperature.

This study marks the first systematic effort to separate the effects of temperature and of electric field across the MIT in a VO$_2$ system. We conclude that electric field is a unique external parameter in driving the MIT: from the $\textit{G}$ vs. $\textit{T}$ curve, the E-MIT creates an interesting correlated metallic phase away from $T_c$, electric field modulates the hysteresis in $\textit{IV}$ curves in ways that are distinct from those of thermal-induced modulations, and E-MIT is created without mediating phase separation for $T << T_c$. In conclusion, we show that an non-equilibrium electric field can create a MIT through an avalanche type process at low temperatures and as $T$ approaches $T_c$, thermal effects become progressively significant. The role of non-equilibrium electric field in modulating charge carries and how microscopically such an avalanche process or discrete steps results require further studies. It is clear that these analyses can also be applied to study systems exhibiting non-equilibrium MIT such as Fe$_3$O$_4$ and several vanadium oxide compounds. 

Acknowledgement: The work is supported by grants NSF DMR 0847324 (AS, SS, TW and SG), NSF DMR 0847169 (LW and SB) and NSF IIP 1311837 (SB). 


\addtolength{\parskip}{\baselineskip}

\end{document}